\documentclass[runningheads]{svmult}

\usepackage{makeidx}   
\usepackage{graphicx}  
\usepackage{subeqnar}  
\usepackage{multicol}  
\usepackage{cropmark} 
\usepackage{physprbb}  
\begin{document}
\title*{The Dark Matter Telescope}
\author{J. Anthony Tyson\inst{1}
\and David M. Wittman\inst{1}
\and J. Roger P. Angel\inst{2}}
\institute{Bell Labs, Lucent Technologies, Murray Hill, NJ 07979, USA
\and Steward Observatory, University of Arizona, Tucson, Arizona 85721, USA}

\authorrunning{Tyson et al.}

\maketitle              

\begin{abstract}  
Weak gravitational lensing enables direct reconstruction of dark
matter maps over cosmologically significant volumes.  This research is
currently telescope-limited.  The Dark Matter Telescope (DMT) is a
proposed 8.4 m telescope with a 3$^{\circ}$ field of view, with an
etendue of 260 m$^2$ deg$^2$, ten times greater than any other
current or planned telescope.  With its large etendue and dedicated
observational mode, the DMT fills a nearly unexplored region of
parameter space and enables projects that would take decades on
current facilities.  The DMT will be able to reach 10$\sigma$ limiting
magnitudes of 27-28 magnitude in the wavelength range $.3 - 1\mu$m
over a 7 deg$^2$ field in 3 nights of dark time.  Here we review its
unique weak lensing cosmology capabilities and the design that enables
those capabilities.
\end{abstract}

\section{Breaking Degeneracies, Testing Foundations}

Direct information on cosmology -- and thus on the early history of
the universe --  can in principle be obtained by measuring the spectrum 
of mass as it evolves with cosmic time.  
The current set of cosmological models (cf. Zaldarriaga etal 1997, Hu
1998, Turner \& Tyson 1999), has over ten free parameters. Each type of
observational test encounters 
degeneracies among the parameters that cannot be resolved without
additional information. Measuring the mass distribution at redshifts 0.2--1
breaks this degeneracy because it is sensitive to different parameter
combinations than is the cosmic microwave background (CMB).
More importantly, such observations and their concordance with the CMB
results will provide an independent test of the validity of the entire
theoretical framework.

From work spanning more than five decades, it has become
apparent that light and mass are not identically distributed.  It is
now possible to discover mass concentrations that are not evident in
the light distribution. Ultimately, such studies will define the
distribution of mass on a variety of scales. What do the inferred dark
mass concentrations imply for cosmological simulations that are
normalized to the number density of luminous mass concentrations?
Such questions will not be addressed by spectroscopic surveys, even
those as large as the Sloan survey.

The only way to directly ``weigh'' the mass distribution is through
weak gravitational lensing.  In its simplest form, gravitational lens
distortions of the distant galaxies enables a reconstruction of the
projected mass density map for the intervening lens.  But one can do
even better: photometric redshifts enable us to slice
the projected sky in redshift bins.  By obtaining weak lensing maps
for sources at a variety of redshifts, we could obtain a
three-dimensional mass map of the universe back to half its current
age. This is called cosmic weak gravitational lens mass tomography.
In addition to enabling a measurement of the geometry of the
universe, a weak lensing survey will also lead to an understanding of
the relationship between mass and light on galactic and cosmological
scales. 

Because structure on the scale of $\sim$ 120 Mpc exists, only a survey
that samples mass in volumes significantly larger than 120 Mpc on a side will
provide a definitive, representative measurement of the distribution
of mass. Moreover, a form of dark energy -- quintessence -- should itself
clump on several hundred Mpc scales, so that a direct mass survey covering
tens of degrees (for mass-energy at redshift 0.5) would probe this.
The metric size of the mass structures, and the redshift at which they
most powerfully lens 24-27 mag background galaxies, sets the angular
scale for each field. 
In fact, the best way to ensure a fair sampling
of the universe is to study several such large, well-separated fields.

Pilot weak lensing surveys are currently underway, but due to the limitation
of existing facilities and time, such studies can at best cover only
much smaller areas (in fewer colors) than the definitive survey
discussed here.  One requires good angular resolution over a
3$^\circ$ field, coupled with the light gathering power of an 8 m
class mirror.  However, the 8 m class telescopes now coming online
fall far short of the field-of-view requirement.  The DMT could reach
the required depth of 29.5--28 mag arcsec$^{-2}$ throughout the wavelength
range 0.3-1 micron (driven by the need for color redshift resolution) 
over a 1000 deg$^2$ area using half the dark nights over five
years, whereas such a survey on existing 8 m telescopes would take
half a century.

The various direct observational tests of cosmology (weak lensing, CMB
anisotropy, and SNe) separately and in combinations can remove
model degeneracies or uncover model failures.  For example, a 1000
deg$^2$ weak lensing project without any photometric redshift
information can only determine $\Omega_{\rm matter}$ to $\pm 0.3$ and
$\Omega_{\Lambda}$ to about $\pm 0.5$, because these parameters affect
the formation of structure in compensating ways.  Including
photometric redshift information breaks this degeneracy through a
measurement of the redshift evolution of structure; errors on the
parameter can in principle be reduced to $\pm 0.02$ and $\pm 0.04$ for
$\Omega_{\rm matter}$ and $\Omega_{\Lambda}$ respectively.  The
degeneracy can alternately be broken by CMB anisotropy observations on
angular scales of less than 1 deg.  Recently, the BOOMERANG experiment
results were announced, showing that the universe appears to be flat
to within ten percent.  Assuming there are no difficulties due to
scattering from the reionization epoch, even higher accuracy CMB observations
over the whole sky are coming in several years (MAP satellite), and even 
more cleanly later with the Planck satellite data.  The
combination of weak lensing ({\it with} photometric redshifts) and CMB
data provides a sharp consistency test for the theory.  In addition
the combination would constrain most of the
parameters ({\it e.g.} Turner \& Tyson, 1999) of current theories
substantially better than either alone.

Taken together with the weak lensing survey and upcoming MAP CMB
anisotropy results, a program of SN Ia observations can put strong
constraints on the equation of state of the contributions to the
mass-energy of the universe, which would easily discriminate between
most of the current contenders: constant vacuum energy density (dark
energy) vs. variable dark energy (quint\-essence) vs. topological
defects of various kinds. Next-generation deep wide-field SN photometric
surveys, covering a range of redshifts within the same calibrated survey, 
will make a significant advance.
The Dark Matter Telescope will discover and follow
thousands of SNe per year. Accurate multi-color photometry on 3000 high redshift 
SNe per year will be obtained in the deep survey mode, and 200,000 moderate
redshift SNe will be found in the wide-deep mode, per year.
Finally, including the results on
large-scale structure as reflected in the luminous local baryonic
component ({\it e.g.} Sloan survey) will yield a complete picture of the
development of the rich structure we see around us.

\section{Cosmology with Weak Lensing}

\begin{figure}
\begin{center}
\includegraphics[width=1.0\textwidth]{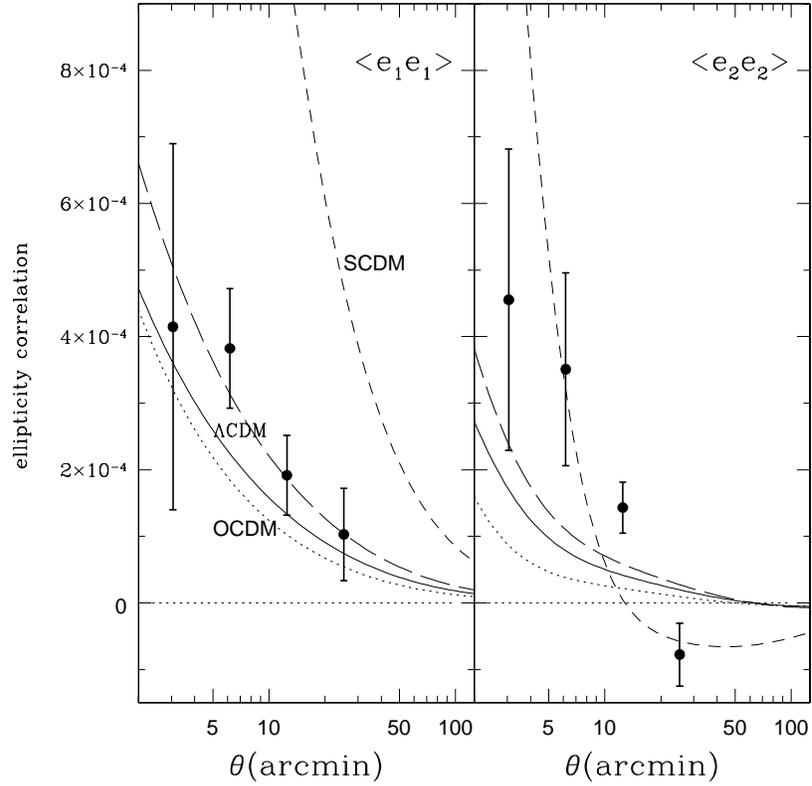}
\end{center}
\caption{The cosmic shear detection of Wittman {\it et al.} (2000), showing
the correlations of $e_1 = e \cos(2\beta)$ and $e_2 = e
\sin(2\beta)$ (where $e$ is the ellipticity and $\beta$ is the
position angle) for pairs of distant galaxies as a function of angular
separation $\theta$.  The predicted correlations depend on the redshift of the
source galaxies and on the cosmological model.  The measured
correlations are plotted here with the predictions of three
cosmologies for their best estimate of the source redshift
distribution: $\Lambda$CDM (solid line) SCDM (short-dash), and OCDM
(dotted). The long-dash line shows the effect of a 20\% error in the
mean source redshift for $\Lambda$CDM.  A cosmological model must match both
autocorrelations; SCDM is ruled out at many sigma by $\langle e_1e_1
\rangle$, while $\Lambda$CDM and OCDM match $\langle e_1e_1 \rangle$ very well and are consistent
with $\langle e_2e_2 \rangle$ at the 3-sigma level.  The
cross-correlation $\langle e_1e_2 \rangle$ (not shown) is consistent
with zero, as expected in the absence of systematic error.  The Deep
Lens Survey now in progress will provide a much stricter test of
cosmological models, or suggest the need for new models.
As shown in the next two figures, the 8.4 m DMT deep weak lensing data will lead
to precision measurements
of the mass spectrum and will tightly constrain several cosmological parameters,
independent of the SN observations.}
\label{fig-ocf}
\end{figure}

For more than thirty years, the search for cosmic shear, or weak
gravitational lensing by large-scale structure, was stymied by
limitations in instruments ({\it e.g.} Kristian 1967; Valdes, Tyson \&
Jarvis 1983; Mould {\it et al.} 1994).  Advances in instruments (large
mosaics of sensitive and linear CCDs, coupled with better telescope
image quality) and data reduction techniques (cancellation of
point-spread function anisotropy) stimulated the first detections of
cosmic shear (Wittman {\it et al.} 2000; van Waerbeke {\it et al.}
2000; Bacon {\it et al.} 2000; Kaiser {\it et al.} 2000).
Figure~\ref{fig-ocf} illustrates the detection of Wittman {\it et
al.} and the predictions of three cosmological models.  The four
papers agree roughly with each other and with the predictions of
$\Lambda$CDM, but the error bars are large; only SCDM can be ruled out
on the basis of the current weak lensing measurements alone.
However, when these preliminary results of weak lensing cosmic shear
are combined with the latest cosmic microwave background determinations of 
a nearly geometrically flat universe, the 2$\sigma$ region of concordance suggests
that $0.25 < \Omega_m < 0.5$ and that the cosmological constant is non-zero and 
in the range $0.4 < \Omega_{\Lambda} < 0.8$. A dramatic increase in accuracy for this
cosmic shear measurement will occur in just a few years.

The next generation of cosmic shear measurements is already underway
and will provide smaller errors over larger angular scales.  The
Deep Lens Survey (http://dls.bell-labs.com), for example, will cover
seven 2$^\circ$ square fields, while the first-generation measurements
used 40' fields at best.  At large scales, cosmic variance is expected
to be the dominant source of error, so covering a number of different
fields is critical.  This project has been granted a large amount of
telescope time (86 nights on 4 m telescopes), reflecting the
increasing importance of cosmic shear measurements, but that time is
stretched over five years.  Several other groups have surveys
underway, and they all seem to be telescope-time limited.

\begin{figure}
\begin{center}
\includegraphics[width=.8\textwidth]{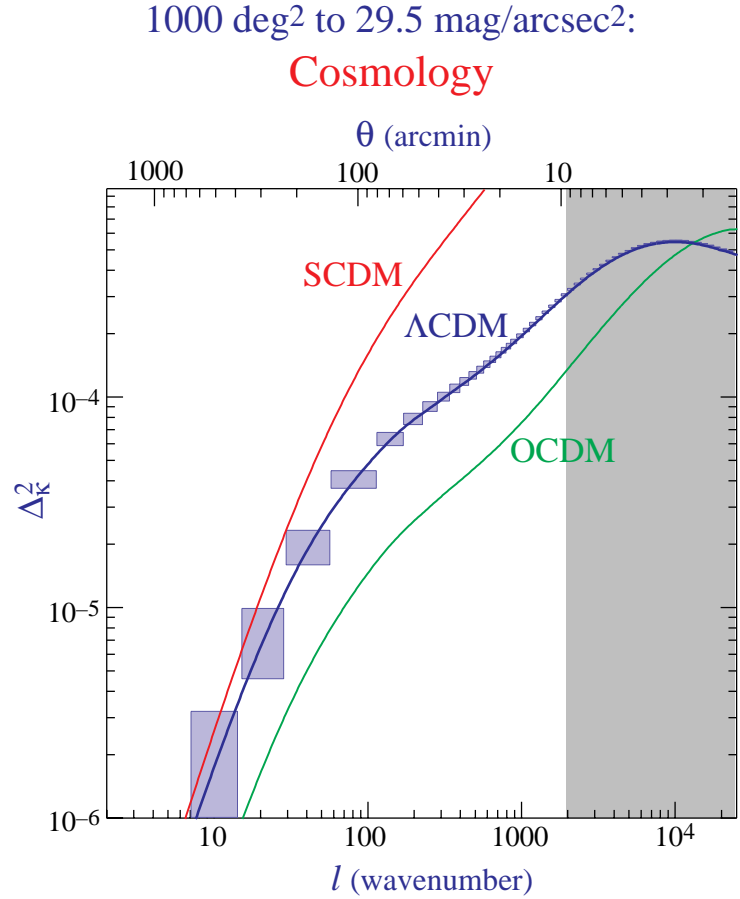}
\end{center}
\caption{Projected errors in the
convergence power spectrum from a 1000 square degree survey with the Dark
Matter Telescope, shown attached to the predictions of $\Lambda$CDM.
The shaded region on the right is where current theory for nonlinear
evolution is unreliable.  Courtesy W. Hu.}
\label{fig-dmterrors}
\end{figure}

\begin{figure}
\begin{center}
\includegraphics[width=.8\textwidth]{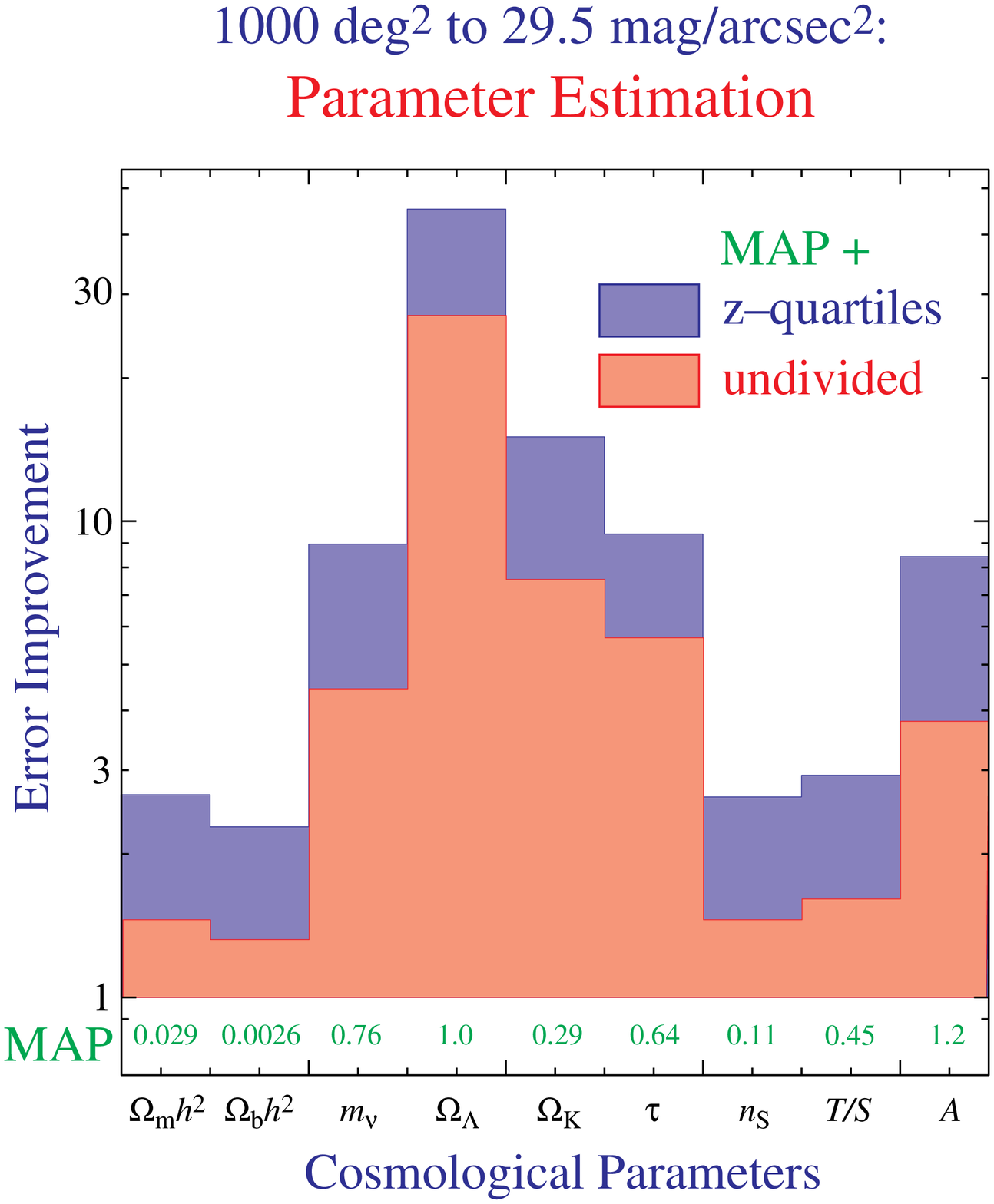}
\end{center}
\caption{Combining the results of the 1000 square degree survey
with MAP data breaks degeneracies and pinpoints several cosmological
parameters to much greater accuracy. The numbers just above the
horizontal axis are the estimated errors of MAP alone.  Courtesy
W. Hu.}
\label{fig-dmtparams}
\end{figure}

In most physics experiments, one has the ability to repeat
measurements while chopping possible sources of error.  Although
astronomers are limited in this respect due to the great distances
of their sources, they are also limited by a lack of telescope time.
Weak lensers could (indeed,
should) repeat their measurement in a different area of the sky, at 
different wavelengths, under different atmospheric
conditions, with different exposure times, and so on; but they normally do not,
simply because of insufficient telescope time.  Any hope of doing
precision cosmology rests on semi-dedicated facilities which can
repeat the measurement numerous times under different conditions, chopping
sources of systematic error on the relevant timescales.  
The design of such a facility is driven by the need to survey large
amounts of sky fairly rapidly, and in the next section we introduce
such a design in some detail.  Figure~\ref{fig-dmterrors} shows how
accurately a weak lensing survey on such a telescope would measure the
shear power spectrum, and Figure~\ref{fig-dmtparams} demonstrates how
such a survey, when combined with data with NASA's MAP mission, breaks
degeneracies on numerous cosmological parameters.

\section{Telescope Design}

\subsection{Optics}
In given integration time, the size of field larger than
$\Omega$ that can be explored to given depth is directly proportional
to the figure of merit $A\Omega\eta/d\Omega$, where A is the
collecting area, $\Omega$ the solid angle of the field of view, $\eta$
the detector quantum efficiency and d$\Omega$ the solid angle of the
seeing-limited image.  Today's 8 m class telescopes are superb at
optimizing all of these factors except $\Omega$.  Conventional
designs, including Schmidt telescopes and other corrected systems
based on one or two mirrors, are incapable of wide fields at the fast
focal ratios required to match detector requirements and minimize
overall cost.

Fundamental to any design are the image size and detector pixel size,
which set the focal length.  At good sites, the atmosphere will
deliver 0.5 arcsec images, while CCDs with 13--15$\mu m$ pixels are
likely to provide enough full well capacity.  For Nyquist sampling of
0.5 arcsec images, a plate scale of 50--60 $\mu m$ arcsec$^{-2}$ is
required.  This implies a focal length of 10--12 m, or a speed of up
to f/1.25 for an 8m primary.  Conventional designs, including Schmidt
telescopes and other corrected systems based on one or two mirrors,
are incapable of wide fields at so fast a focus.  

However, three-mirror designs with this capability were first explored
by Paul (1935).  He gave a design with a parabolic primary, convex
spherical secondary and a concave spherical tertiary of equal but
opposite curvature.  The image is formed midway between secondary and
tertiary, with good correction over a wide field.  One can think of
the design as a reflective Schmidt telescope used as a corrector for a
large afocal Cassegrain telescope.  The secondary, located at the
center of curvature of the tertiary, has added correction for
spherical aberration similar to a reflecting Schmidt plate.  A
telescope of this type was built by McGraw {\it et al.} (1982), using
a 1.8 m parabolic primary at f/2.2, giving images no more than 0.2
arcsec rms diameter at the edge of the field.
In addition, an
all-reflective design is critical for reducing the scattered light
problems that accompany a large field of view.

\begin{figure}
\begin{center}
\includegraphics[width=.8\textwidth,angle=-80]{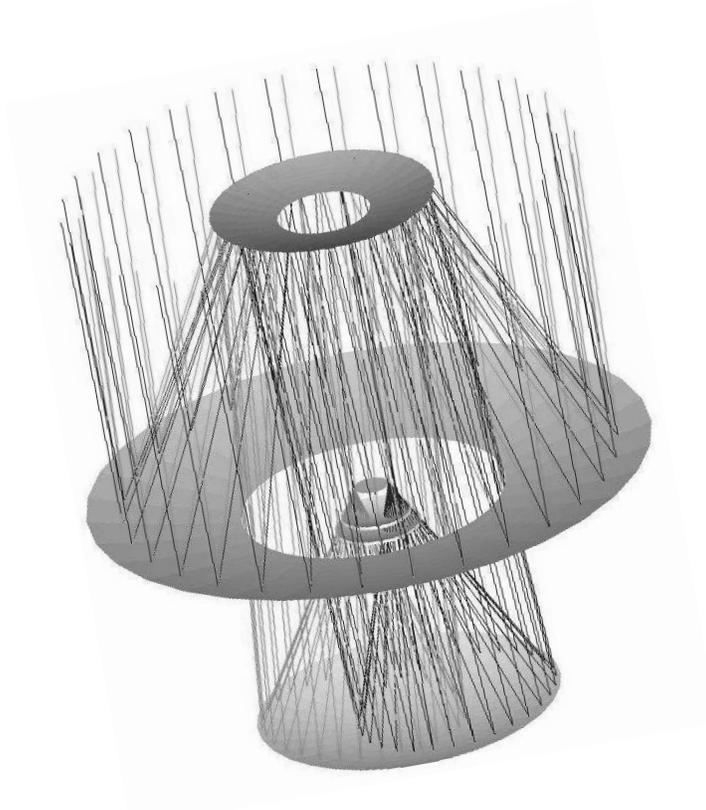}
\end{center}
\caption{DMT optical layout with rays covering a 3 deg field.}
\label{dmt}
\end{figure}

Angel {\it et al.} (2000) used this design as the starting point for
exploring more general three-mirror systems using computerized
optimization.  An optimum design balances obscuration and field of
view to maximize etendue or A$\Omega$ product for given primary
aperture.  They found that very well corrected fields of up to
3$^\circ$ diameter could be formed at f/1.25 using the three mirrors
alone.  The mirrors are arranged so the light from the secondary
passes through a half-diameter hole in the primary to a near-spherical
tertiary behind, and the light comes to a focus near the primary
vertex (Figure~\ref{dmt}).  To minimize the secondary obscuration, the
primary focal ratio was held at f/1.0.  Detector obscuration was
minimized by making the primary and secondary together afocal.  When
the 3$^\circ$ field is completely baffled against stray sky light
illumination the total obscuration and vignetting is 26\% at the field
center, rising to 38\% at the field edge, making an effective aperture
of 6.9 m.

\subsection{Sky Baffling}
Through most of the telescope's spectral range, up to 1.8$\mu$m, its
sensitivity is limited by photon noise from optical emission by the
atmosphere.  To prevent additional skylight from reaching the focal
surface indirectly, two black conical baffles will be used.  The
first, 4 m in diameter, extends 0.5 m below the secondary mirror.
The second, 3.8 m diameter, rises 1.1 m above the primary hole. These
two give complete sky baffling out to the full 3$^{\circ}$ field.

\subsection{Mechanical Considerations}
\begin{figure}
\begin{center}
\center{\includegraphics[width=.8\textwidth]{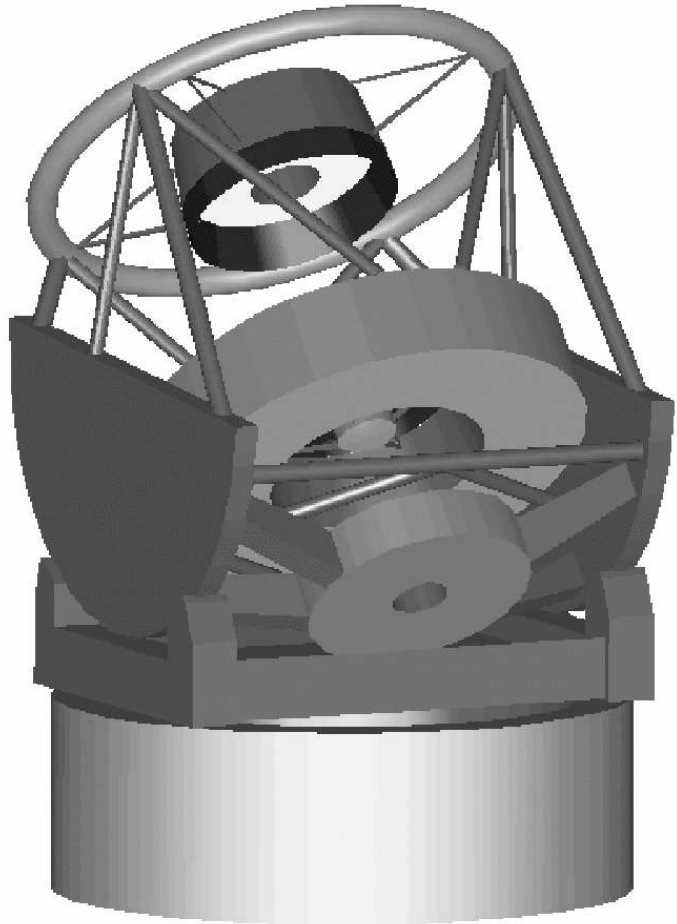}}
\end{center}
\caption{The proposed DMT rigid, fast-slewing mount. The three mirrors are held between large C
rings turning on a flat azimuth platform. The relatively short telescope, given its
8.4 m primary mirror, translates to a low cost dome.}\label{angeljrp4.EPS}
\label{fig-dmt}
\end{figure}

The optical assembly for an 8.4 m primary is short, only 9 meters
between the secondary and tertiary, with the primary and camera set
midway between.  This configuration is advantageous both for making an
agile telescope and an inexpensive enclosure.
All three large mirrors would be stiffly mounted between two C rings,
supported on a compact azimuth frame that transmits loads directly to
a large diameter pier (Figure~\ref{fig-dmt}). The stiffness of the drives
gains by the square of the C ring radius, and we find that the large
semicircular rings shown, 11 m diameter, coupled with the relatively
small moment of inertia lead to excellent tracking and stability.
Repointing by 3$^{\circ}$ in a time as short as 5 seconds is
realistic.  The telescope enclosure will be smaller and hence less
expensive than for a standard 8 m telescope, because of the small
turning radius.

\subsection{Camera}

CCD detector arrays are now a rather mature technology, and there is
little doubt that a mosaic to cover the full 55 cm circular focal
surface can be built for acceptable cost.  Covering the 3$^\circ$
diameter field of view with 0.25 arcsec pixels (to provide critical
sampling in the best seeing) requires 1.4 Gpix.  The individual CCDs
will be small, 1k or perhaps 2k, so that the the circular field and
the curvature of the focal plane can be matched precisely.  This
multiplexing also allows for fast readout, which will be critical for
efficiency, as exposure times must be short, $\sim 30$ s, to avoid
saturation.  The CCD array will be 55 cm in diameter, in a dewar with
a robust fused silica vacuum window.  This window (plus any filter) is
the only refractive element in the system, so that the scattered light
resulting from bright stars in the huge field will be minimized.  The
aberrations that would be introduced by this element are balanced by
adjusting the prescriptions of the three mirrors.

A significant issue for CCDs is the control of charge blooming from
bright stars.  Deep wells, anti-blooming measures and the use of many
smaller devices will all be important control measures. We shall
suppose that detectors with 13 $\mu$m pixels (0.25 arcsec) are
used. These should be manufacturable with deep wells; already full
well as high as 150,000 electrons has been demonstrated for 8x8 $\mu$m
pixels.  In the near future, 13 $\mu$m pixels could be
optimized for still greater capacity.  Anti-blooming capabilities can
be incorporated in the detectors, which may reduce the full well
capacity.  Alternatively, anti-blooming clocking schemes may be used
during integrations. To avoid uncontrolled blooming from the brightest
stars, (a few really bright stars will be inevitable in a 3$^{\circ}$
field), a large number of smaller format devices may be preferred.
These could be as small as 1024 pixels square, in which case 1300
devices would be needed to tile the 55 cm diameter focal plane.  We
find that the read time should be no more than 5 seconds,
requiring a realistic 200 kHz pixel rate to read each 1024 square
device with a single amplifier.

As a way to minimize the gaps between the individual CCDs we are
presently exploring detector packaging techniques which will allow the
use of true, 4-side buttable devices using semiconductor industry
standard packaging technologies. This development would limit the
inter-device gaps to the non-imaging silicon of each detector, which
is dominated by clock busses, amplifiers, and, most significantly, I/O
bonding pads.  With the continued industry-wide trend toward smaller
I/O structures, it is not unreasonable to expect 50$\mu$m bond pads to
be sufficient for future CCDs.  If we therefore assume uniform gaps of
100 $\mu$m around each CCD, a fill factor of 96\% can be obtained.

Cooling requirements are not severe for the CCD mosaic.  The criterion
is that the dark rate be less than the sky photon rate in the darkest
filters.  We estimate that even for the 360-nm band or the narrowest 3\%
filter that photon rates will be $> 1.5 e^{-}$/pixel/sec.  With an MPP
device, dark rates less than this can be achieved at a device
temperature of about -15 C.  It follows also that in the worst case of
a 20 second exposure in a dark band a read noise of $\sim$ 4 e$^-$ rms
will be acceptable.

While certainly a very large number of devices are required for this
project, the CCDs themselves could be manufactured today.  The DC
shorts yield of several fabrication lines is now over 50\%.  If 50\%
of these unshorted thick CCDs are of astronomical quality, a mature
lot run will yield about 25\% useable devices from 6" silicon wafers
at a typical fab facility.  Assuming a 25-50\% thinning and packaging
yield, the final thinned device yield would be about 5-10\% of the
starting lot.  This would then require about $\sim$ 200 wafers to be
fabricated with 100 devices per wafer, after one or two engineering
and test lots.  Cryogenic DC and AC wafer probing will allow rapid
feedback to the fabricator on device quality and yield.

\subsection{Comparison with some Existing and Proposed Imaging Telescopes} 

The etendue at the 3$^{\circ}$ focal plane of the Dark Matter
Telescope is 260 m$^2$ deg$^2$.  The most powerful imaging
telescope currently in operation is the Sloan Digital Sky Survey.  It
has a modified Cassegrain system with 2.5 m aperture and a 3$^{\circ}$
field at f/5.  Comparing the 8.4 m telescope with the SDSS, and
allowing also for its increased pixel sampling and resolution, the
advantage in figure of merit is by a factor of close to 100. The wide
field optical cameras to be used with larger telescopes, such as
Subaru's Suprime and MMT's Megacam have etendues which are not
substantially larger than the SDSS, in the range 5 - 10
m$^2$~deg$^2$.
This telescope will provide a capability that is completely
beyond any existing telescope and uncovers a region of parameter
space orders of magnitude beyond current limitations. 
The Dark Matter Telescope design and capabilities are discussed in more
depth on the website http://dmtelescope.org.

\section{Sensitivity and Observing Strategies}

We have estimated imaging sensitivity by
scaling the 10$\sigma$ point source magnitude limits in the
Johnson photometric bands U - I (0.35 - 0.90 $\mu$m) derived ab-initio
by Angel {\it et al.} (1999) to an 8.4 m telescope.
These included the blurring effects of atmospheric seeing and
dispersion at 45$^{\circ}$ elevation.  These limiting magnitudes are listed in 
Table 1. For this 8.4 m telescope, the red limiting magnitude for a 20 sec
exposure is just sufficient for an all-available-sky survey (sharing time with
the deep survey) to detect and confirm 90\% of the near-Earth objects in a decade. 

\begin{table}
\caption{10$\sigma$ limits for the 8.4 m telescope}\label{tbl-2}
\begin{center}
\renewcommand{\arraystretch}{1.2}
\setlength\tabcolsep{5pt}
\begin{tabular}{|c|c|c|}\hline
Band \hspace{.125in} $\lambda$(mm) & 20 second exposure & 9 hour exposure\\ \hline
\hspace{0.05in}U \hspace{.17in}0.365 & 22.8 & 26.8\\
B \hspace{.125in} 0.44 & 23.8  & 27.8\\
V \hspace{.125in} 0.55 & 23.9  &27.9\\
R \hspace{.125in} 0.70 & 23.6  & 27.6\\
I \hspace{.16in} 0.90 &  22.8  & 26.8\\ \hline

\end{tabular}
\end{center}
\end{table}

Because the exposure time is limited to about 30 s to avoid CCD
saturation, the deep imaging needed for lensing work must be
accumulated over many exposures.  This observing mode also makes the
data useful for many other projects which require short exposures.
The search for near-Earth objects and outer solar system objects will
benefit greatly, and a vast amount of parameter space will be opened
up in the search for GRB counterparts and previously unknown types of
optical transients.  A thorough search for high-redshift supernovae (which
break the degeneracy in cosmological parameters in a different way
than weak lensing) can also be carried out with this data.  The
traditional mode of observing must give way to a dedicated program,
with the community sharing the data.

The depth of the 20 second exposures suggests one possible observing
mode: repeatedly surveying the entire observable sky to 24th magnitude
in 3-4 nights.
Assuming the readout is accomplished while the telescope is repointed
in 5 seconds, 144 exposures an hour could be obtained in clear
weather, and the 3000-exposure survey would take about 22 hours, i.e.
3-4 nights.  Image differencing between successive surveys will reveal
variable and moving objects, and image summation over many such
surveys will provide deeper images. About half the
time would also be dedicated to extremely deep multi-band images
over smaller regions, such as the several thousand square degrees required
for the wide-deep cosmology projects described above.
The last column in Table 1 gives the limiting
magnitudes for 9 hours of exposure, assuming that sensitivity
increases as the square root of the integration time.  In a week of
clear weather, a single 3$^{\circ}$ field could be observed to the
given depth in each of the five filters, for accurate photometric
redshifts. The high etendue corresponding to the 8.4m aperture permits 
the parallel execution of these two observing modes, completing both
the all-sky search and the ultra-deep projects within a decade.  

The data rate from a 1.4 Gpix camera with 30 s exposures sounds
prodigious---over 1 TB per night---yet such data rates are commonplace
in radio astronomy and particle physics.  Routine image processing on
large images is easily parallelized and ideally suited to clusters of
inexpensive, commodity computers, especially given the parallel nature
of the readout.  Data storage on commodity hardware will also be
feasible by the time of DMT operation.

\section{Summary}
Weak gravitational lensing can break the degeneracies in CMB
anisotropy measurements of cosmological parameters.  To provide
comparable precision, weak lensing needs a dedicated or semi-dedicated
telescope facility.  We propose the Dark Matter Telescope, a 8.4 m
telescope with a 3$^\circ$ field of view and good image quality,
providing an unprecedented figure of merit for deep surveys.
After discussing the design and observing strategy, we
conclude that such a telescope is feasible now, and show its potential
impact on cosmology.

This proposed facility embodies a non-traditional approach to
ground-based optical/IR astronomy: equal emphasis is
placed on the survey products, their unique science
capability and distribution to the community, the
data pipeline, the camera and data system, and the
telescope. As such, this project would be pursued in
a manner similar to those of experimental high-energy physics.
The 8.4 m aperture and 7 deg$^2$ field enable huge advances in other
areas as well: planetary astronomy (Kuiper belt objects and near-Earth
objects), and the transient universe (gamma-ray burster afterglows over
lage volumes, and new classes of objects).
In the ranked projects in the recent NRC AASC Decadal Survey for Astronomy, 
the DMT was named the ``Large-area Synoptic Survey Telescope'' (LSST) to
emphasize its complementary applications and multiple missions.

\end{document}